\documentclass[times,review]{elsarticle}

\usepackage{jcomp}
\usepackage{framed,multirow}

\usepackage{amssymb}
\usepackage{latexsym}
\usepackage{amssymb}
\usepackage{amsmath}
\usepackage{bm}
\usepackage{amsthm}
\usepackage{xcolor}
\usepackage{mathtools}
\usepackage{lineno}
\usepackage{caption}
\usepackage{subcaption}
\usepackage{float}

\usepackage{url}
\usepackage{xcolor}
\definecolor{newcolor}{rgb}{.8,.349,.1}

\newcommand{\fm}{\bm{m}}
\newcommand{\fn}{\bm{n}}
\newcommand{\fx}{\bm{x}}
\newcommand{\fy}{\bm{y}}

\graphicspath{{./plots/}}

\journal{Journal of Computational Physics}

\begin{document}

\verso{B. Dorschner \textit{et al.}}

\begin{frontmatter}

\title{A fast multi-resolution lattice Green's function method for elliptic difference equations}

%

\author[1]{Benedikt Dorschner\corref{mycorrespondingauthor}}
\ead{bdorschn@caltech.edu}
\cortext[mycorrespondingauthor]{Corresponding author}
\author[1]{ Ke Yu}
\author[1]{ Gianmarco Mengaldo}
\author[1]{ Tim Colonius}
\address[1]{California Institute of Technology, 1200 E California Blvd, CA 91125, Pasadena, U.S.A.}

\begin{abstract}

We propose a mesh refinement technique for solving elliptic difference
equations on unbounded domains based on the fast lattice Green’s function
(FLGF) method.  The FLGF method exploits the regularity of the Cartesian mesh
and uses the fast multipole method in conjunction with fast Fourier transforms
to yield linear complexity and decrease time-to-solution.  We extend this
method to a multi-resolution scheme and allow for locally refined Cartesian
blocks embedded in the computational domain.  Appropriately chosen
interpolation and regularization operators retain consistency between the
discrete Laplace operator and its inverse on the unbounded domain. Second-order
accuracy and linear complexity are maintained, while significantly reducing the
number of degrees of freedom and hence the computational cost.

\end{abstract}

\begin{keyword}
Elliptic difference equation \sep Poisson problem  \sep Lattice Green's
function\sep Fast multipole method \sep Mesh refinement

\end{keyword}

\end{frontmatter}


\section{Introduction}
\label{section:0}

Elliptic partial differential equations require fast numerical solvers scalable up to trillions of
unknowns. In this contribution, we will focus on elliptic difference equations
on unbounded domains arising in numerous applications including the incompressible Navier-Stokes equations
(NSE), quantum mechanics, random walks, and plasma physics.
\cite{economou1983green,mccrea1940xxii,kulikovsky1997positive}.  
%
In particular, we will use the Poisson equation as a representative example of
these.  Common, scalable methods to solve these equations include the fast
Fourier transform (FFT) \cite{cooley1965algorithm}, the fast multipole method
(FMM) \cite{greengard1987fast}, multigrid methods and its variants (algebraic
(AMG) and geometric (GMG)) \cite{ruge1987algebraic}, domain decomposition
methods \cite{smith2004domain} as well as wavelet transforms
\cite{schneider2010wavelet}.  A recent comparison of FFT, FMM as well as AMG
and MGM can be found in \cite{gholami2016fft}, where the Poisson equation was
solved in the unit cube with periodic boundary conditions and scaled up to 600
billion  degrees of freedom (DoF) and $220,379$ cores.  The authors conclude
that while FFT is the method of choice for smooth source functions that require
uniform resolution, it is outperformed by the FMM for localized source
distributions where one can take advantage of nonuniform grids. In addition,
state-of-the-art algebraic multigrid solvers were found to be an order of
magnitude slower than FFT, GMG or FMM. Note that while the authors compare
these methods for periodic boundary conditions only, we focus on
free-space boundary conditions in the following. In this case, it is expected
that the efficiency between both FFT and multigrid method will decrease due to
the difficulties of satisfying free-space boundary with the appropriate
Dirichlet boundary conditions or zero-padding, respectively.

A fast lattice Green's function method (FLGF) has recently been
proposed in \cite{liska2014} and shown extraordinary efficiency in solving the
class of elliptic difference equations in unbounded domains. 
The FLGF method is based on a efficient convolution of the source distribution and 
its fundamental solution via FMM, the Green's function. 
In contrast to most existing FMM approaches, the FLGF method exploits the regularity
of a Cartesian mesh, and uses a kernel-independent, interpolation-based fast
multipole method (FMM) to retain linear computational complexity inherent to
FMM. A similar approach has been proposed in \cite{gillman2014fast} for two 
dimensional infinite lattices using skeleton points (see also \cite{gillman2010fast,martinsson2002fast}). 
Additionally, the FLGF exploits fast Fourier transforms (FFT) to further reduce the
computational costs compared to conventional solvers.  
It 
%
%
has 
been successfully applied to solve the incompressible Navier-Stokes equations
using a finite volume approach \cite{liska2016}.  In addition, accurate
simulations of external aerodynamics of complex geometries were enabled by
coupling it with the immersed boundary method (IB-LGF)\cite{liska2017}.
Owing to the geometrical flexibility of the FMM and the implied free-space
boundary of the Lattice Green's functions (LGF), the solver allows for
adaptive, possibly disjoint meshes that limit the computational domain to a set
of blocks with non-negligible vorticity only, with further computational
savings associated with the typical compactness of this region.
This is in contrast to most common approaches, which employ spatially truncated
domains with approximate free-space boundaries. These approximations introduce
blockage errors, which affect accuracy and may even change the dynamics of the
flow \cite{Tsynkov1998,Colonius2004,pradeep_hussain_2004}.  Thus, large
computational domains in combination with stretched grids
\cite{Taira2007,Yun2006a,Wang2011a}, local refinement
\cite{roma1999adaptive,griffith2007adaptive} and far-field approximations
\cite{Colonius2008} are required to limit the influence of the approximate
free-space boundary condition. 

A significant limitation of the FLGF approach, however, is uniform
resolution. This limits its ability to reach competitive time-to-solution
requirements, commonly required by demanding academic or industrial
applications. In fact, while uniform Cartesian meshes can significantly
decrease the cost per degree of freedom (DoF), the total number of DoF can be
prohibitive for strongly anisotropic or inhomogeneous problems with localized
source regions. This issue is particularly prominent for, e.g., high Reynolds
number flows or problems where the range of scales is large. 
The use of uniform meshes for this set of problems requires a resolution that
is comparable to the smallest scales in the problem, although these might be
confined to a very small region of the computational domain, thereby yielding
an unnecessarily high number of DoF.
%
To overcome this limitation, we propose here a multi-resolution extension of
the FLGF method. 
Most popular multi-resolution schemes include multigrid  methods as well as
FMM, which inherently support local mesh refinement and have been successfully
implemented in a variety of publicly available software packages (see, e.g.:
\cite{lashuk2009massivelyFMM,ying2004kernel,sundar2012parallelAMG} or \cite{dubey2014survey} for a
review).  However, our goal here is to retain
the favorable efficiency of the FLGF method compared to the classical FMM as
well as multigrid methods but reduce the number of DoFs by a proposing
block-structured mesh refinement algorithm for the FLGF method.


The paper is organized is follows:
In section \ref{section:1}, we start by reviewing the fast lattice Green's
function method in some detail to facilitate further discussion.  Subsequently,
the proposed multi-resolution scheme is presented in section \ref{section:2}. In addition, 
we provide an analysis, discussion and a convergence study of the refinement scheme 
on a fluid dynamics test problem.

\section{Fast lattice Green's function method}
\label{section:1}
The lattice Green's function method \cite{liska2014} solves inhomogeneous,
linear, constant-coefficient difference equations, defined on unbounded
Cartesian grids, by convolution of the fundamental solution of the discrete
operator, so-called lattice Green's functions (LGF), with the equation's source
term. 
LGFs can be obtained from Fourier integrals, and its asymptotic expansions can
be used to facilitate numerical or analytical evaluation.  The formally
infinite mesh may however be truncated such that only cells with non-negligible
source are retained. Going by example, for block-structured meshes this allows
us to adapt the computational domain by simply adding or removing the
corresponding blocks.

Here, we will focus on elliptic difference equations on unbounded domains,
exemplified by 
the three dimensional Poisson's problem, which reads
\begin{equation}
  \Delta u  (\bm x)=f(\bm x), \quad supp(f)\subseteq \Omega,
\end{equation}
where $\bm x \in \mathbb{R}^3$ and $\Omega$ denotes a bounded domain in
$\mathbb{R}^3$. Its solution or target field $u$  can be obtained by 
convolution of the fundamental solution of the Laplace operator ${\rm{G}}(\bm x)= -1/(4\pi |\bm x|)$ 
with the source field $f(\fx)$ such that
\begin{equation}
u ( \bm{x}) = ({\rm{G}} \ast f)(\fx )=\int_\Omega {\rm{G}}(\fx - \bm \xi) f( \bm \xi) {\rm{d}} \bm{\xi}.
\end{equation}
Correspondingly, the discrete scalar Poisson equation reads
\begin{equation}
  L_{\mathcal{Q}} u  (\bm n)=f(\bm n), \quad supp(f)\subseteq \Omega_h,
\end{equation}
where $u,f \in \mathbb{R}^{\mathcal Q}$, $\Omega_h$ is a bounded domain in
$\mathbb{Z}^3$, $\bm{n}=(n_1,n_2,n_3) \in \mathbb{Z}^3$ and $ \mathcal Q \in \{\mathcal C\}$ denotes cell-centered values.
Its solution is given by the discrete convolution 
\begin{equation}
\label{eq:discretConv}
u ( \bm n) = (G \ast f)(\bm n )=\sum_{\bm m \in \Omega_h} G(\bm n - \bm m) f( \bm m),
\end{equation}
where  the fundamental solution or
LGF of the discrete $7-$pt Laplacian is denoted by $G$. 
An expression for $G(\bm n)$ can be obtained by diagonalizing the Laplace
operator $L_{\mathcal{Q}}$ in Fourier space, inversion and a back-transform (see, e.g.,
\cite{delves2001green,glasser1977extended}). In terms of Fourier integrals this eventually yields
\begin{equation}
\label{eq:GL}
G(\bm{n})=\frac{1}{8 \pi^3 } 
          \int_{[ -\pi, \pi ]^3} 
            \frac{ \exp{(-i \bm{n} \bm{\xi}) } }
                 { 2\cos(\xi_1) + 2\cos(\xi_2) + 2\cos(\xi_3) -6 } 
          \rm{d} \bm{\xi}.
\end{equation}
For evaluation of the LGF it is typically more convenient to rewrite  
Eq.\eqref{eq:GL} as a one dimensional, semi-infinite integral as 
\begin{equation} 
\label{eq:GL_B}
G(\bm{n})= - \int_0^\infty \exp(-6 t) I_{n_1}(2t) I_{n_2}(2t) I_{n_3}(2t) \rm{d}t,
\end{equation} 
where $I_k(t)$ denotes the modified Bessel function of first kind and order $k$.
While Eq.\eqref{eq:GL_B} is 
readily evaluated using an adaptive Gauss-Kronrod quadrature or alike, it is more 
efficient to evaluate the Green's function through its asymptotic expansion in
the far-field \cite{martinsson2002asymptotic,martinsson2002fast,mangad1967asymptotic}, i.e., large $|\fn|$.
In particular, the target field $u$ can then be written as
\begin{equation}
    u(\bm{n}) = u^{\rm{near}} (\bm{n}) + u^{\rm{far}}(\bm{n}) + \varepsilon(\bm{n}),
\end{equation}
where
\begin{align}
    u^{\rm{near}} (\bm{n}) &= \sum_{ \fm \in \Omega_h^{\rm{near}}(\fn) } G(\fn -\fm) f(\fm) \\
    u^{\rm{far}}(\bm{n}) &= \sum_{\fm \in  \Omega_h \setminus \Omega_h^{\rm{near}}(\fn) } A_G^q(\fn -\fm) f(\fm), 
\end{align}
and $\Omega_h^{\rm{near}}$, $\varepsilon(\bm{n})$ are the near field and the
error due to approximating $G(\fn)$ with $A_G^q(\fn)$ in the far field,
respectively.
The $q$-term asymptotic expansion of $G(\fn)$ is defined such that
$G(\fn) =A^q_G( \fn  ) + \mathcal{O}(|\fn|^{-2q-1} )$. For $q=2$ and 
$q=3$ it reads
\begin{equation}
A_G^2 (\fx) = -\frac{1}{4 \pi |\fx| } 
              - \frac{x_1^4+x_2^4+x_3^4 - 3 x_1^2 x_2^2 
                                        - 3 x_1^2 x_3^2 
                                        - 3 x_2^2 x_3^2  }
                     {16 \pi |\fx|^7}.
\end{equation}
and 
\begin{equation}
\begin{split}
A_G^3 (\fx) =& A_G^2 (\fx) +
\frac{1}{128 \pi  \left| x\right| ^{13} } 
\left(
-228 \left(x_2^2 x_3^2 x_1^4+x_2^2 x_3^4 x_1^2+x_2^4 x_3^2 x_1^2\right)+ 
621 \left(x_1^4 x_2^4+x_3^4 x_2^4+x_1^4 x_3^4\right)-  \right. \\ &
244 \left(x_2^2 x_1^6+x_3^2 x_1^6+x_2^6 x_1^2+x_3^6 x_1^2+x_2^2 x_3^6+x_2^6 x_3^2\right)+
\left. 23 \left(x_1^8+x_2^8+x_3^8\right)  \right),
\end{split}
\end{equation}
respectively.
In our implementation, we use a tabulated integration of Eq.\eqref{eq:GL_B} in
the near field for $|\fn| \leq 100$. 
The asymptotic expansion is chosen to ensure
a conservative error bound of the asymptotic expansion compared to the direct
integration of $ | \varepsilon | \lesssim 10^{-12}$, while keeping the 
number of terms to a minimum.
Hence, $q=3$ is used for $100< |\fn| \leq 600$, whereas $q=2$ suffices for $|\fn|>600$.


\subsection{Fast convolutions}
\label{sec:conv}

The direct evaluation of the discrete convolution in Eq.\eqref{eq:discretConv}
is prohibitive for large computational domains as it requires
$\mathcal{O}(N^2)$ work for $N$ degrees of freedom.
The FLGF method \cite{liska2014} on the other hand employs a kernel-independent
interpolation-based fast multipole method (FMM) to compute the discrete
convolutions in conjunction with block-wise FFT convolution.  The FMM achieves
linear complexity $\mathcal{O}(N)$ by exploiting the fact that, for an elliptic
kernel, the solution in the far-field is much smoother than in the near-field.
Thus, a low-rank representation of the kernel is sufficient to accurately
compute the contribution of far-field, while only the near field requires
full-rank representation of the kernel.
A low-rank approximation of a kernel $K(\fx,\fy)$ can be obtained by 
interpolation using coarse grained sampling points $\fx_0,. . .,\fx_{n-1}$. With the generic
interpolation function $\phi(\fx)$, this yields
\begin{equation}
\tilde{K}(\fx,\fy)=\sum_{i=0}^{n-1} \sum_{j=0}^{n-1} \phi(\fx) K(\fx_i, \fy_i) \phi(\fy),
\end{equation}
and the discrete convolution can then be approximated by
\begin{equation}
\label{eq:fmm_conv_combined}
u(\fx_i) \approx 
    \sum_{j=0}^{M-1} \tilde{K} (\fx_i, \fy_j)  f( \fy_j ) 
    = 
    \sum_{j=0}^{M-1} \sum_{p=0}^{n-1} \sum_{q=0}^{n-1} \phi(\fx_i) K(\fx_p, \fy_q) \phi(\fy_j)  f(\fy_j) 
    \quad i=0,.. ., N-1,
\end{equation}
where $N$ is the number of target points and $M$ the number of source points.
The near- and far-field contributions are treated by constructing a
hierarchical decomposition of the domain using an octree structure
$\mathcal{T}$ (quadtree in two dimensions), for which
Eq.\eqref{eq:fmm_conv_combined} is evaluated recursively. The octree is defined
to have a depth $L_B$, where the tree root is assumed to have level
$0$ and the base level $L_B-1$ corresponds to physical domain.
The tree nodes are 
also referred to as octants and octants without children are leaf nodes. 
The set of leafs on level $l$ is indicated by $B_{\rm{Leafs}}^l$.
A distinct feature of the FLGF as proposed in \cite{liska2014} is that each
tree node corresponds to  a region, which is represented by a Cartesian block and 
contains $N_b=n_b^3$ cells. We further denote the $i$-th octant or block at level $l$ by $\mathcal{B}_i^l$
and the set of all octants at level $l$ by $B^l= \bigcup_{i=0}^{N_B^l}  \mathcal{B}_i^l$, where 
$N_B^l$ is the number of octants on level $l$.
The set of children and the parents are denoted by $\mathcal{C}(\mathcal{B}_i^l)$ and 
$\mathcal{P}(\mathcal{B}_i^l)$, respectively.

\begin{figure}[!t]
	\centering					
	\includegraphics[width=0.9\textwidth]{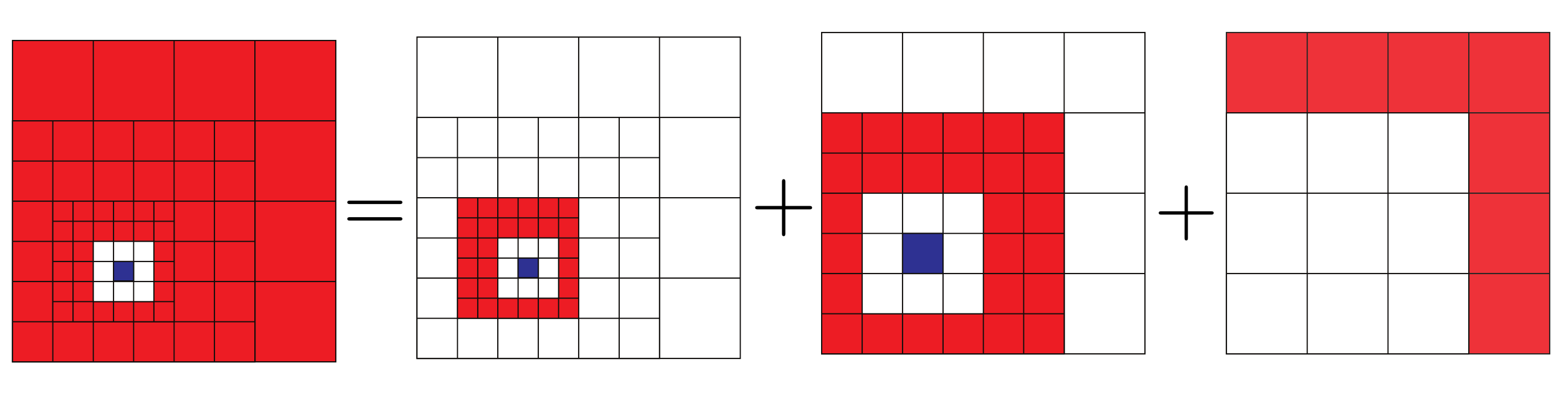}
    \caption{Schematic of the hierarchical domain composition of the far-field
    (red, left) for an octant $\mathcal{B}_i^{L_B-1}$ (blue, left). While the
    near field consist of the nearest neighbors only, the far-field is composed
    out of the set of influence lists for all levels. For level $l$ the
    influence list contains the children of the nearest neighbors of
    $\mathcal{B}_i^{l}$'s parent, which are not contained in the near field,
    i.e., are well separated.}
	\label{fig:fmm_regions}
\end{figure}

Thus, for a given target field $u_i^{L_{B}-1}$, defined on the octant
$\mathcal{B}_i^{L_{B}-1}$ on level $L_{B}-1$, the near-field contribution consists of the
interaction, i.e., convolution, with region $\mathcal{N}(\mathcal{B}_i^{L_{B}-1})$, containing the octant
itself and the nearest neighbor octants on the finest tree level $L_{B}-1$. 
The far-field contributions are then evaluated recursively for the levels $l=L_{B}-1,\dots,
0$ and are defined as the convolution with octants in the influence region
$\mathcal{I} (\mathcal{B}_i^l) = \{ \hat{\mathcal{B}}_i^l \in \mathcal{F}(\mathcal{B}_i^l) \setminus \mathcal{F}(\mathcal{B}_i^{l-1}) \}$, 
which includes the well-separated octants, i.e. 
$\mathcal{F} (\mathcal{B}_i^l) = \bigcup_{l=0}^{L_{B}-1} {B}^l \setminus \mathcal{N}(\mathcal{B}_i^l)$,
but excludes the regions well-separated from its parents $\mathcal{F}(\mathcal{P}(\mathcal{B}_i^{l}))$.
Schematically, the domain decomposition in near and far-field regions is
depicted in figure \ref{fig:fmm_regions}.

Note that using Cartesian blocks as octants allows for the convolution between each
block and its influence list to be computed by a block-wise FFT-based
convolution. There, the convolution is converted to a complex Hadamard product
in Fourier space. As this is a circular convolution, the original block needs
to be zero-padded. FFT-based convolution is a standard technique and we
refrain from reporting details in the interest of brevity but refer the reader to \cite{liska2014}.
Compared to a direct summation as indicated in Eq.\eqref{eq:discretConv} the
block-wise FFT decreases computational complexity from $\mathcal{O}(N_b^2)$ to $\mathcal{O}( N_b \log N_b)$
for each block-convolution. 
It should be clear from the above that given a union of source $B_s=\bigcup_{i=0}^{N_s-1} \mathcal{B}_{s,i}$ and target
blocks $B_t=\bigcup_{i=0}^{N_t-1} \mathcal{B}_{t,i}$ the convolution can be evaluated as the sum of the individual
convolutions as 
\begin{equation}
 u_i =\sum_{j \in B_s} {\rm{conv}}(G_{j-i}, f_j ), \quad \text{for } i=0,. . .,N_t-1,
\end{equation}
where the convolution operator is denoted as $\rm{conv}$ and $G_{j-i}$ is the vector
containing the unique values of $G(\fm-\fn)$ evaluated on the grid points $\fm$ and $\fn$ of the 
blocks $\mathcal{B}_{t,j}$ and $\mathcal{B}_{s,i}$, respectively. In the case of 
FLGF this convolution is evaluated using FFT. 
The additional advantage of such a block-structured FMM approach is that 
regions with negligible source can be removed entirely, yielding an adaptive and 
possibly disjoint mesh.

With these definitions and the corresponding tree structure, the evaluation of
Eq.\eqref{eq:fmm_conv_combined} can be split into three consecutive steps.
First, we iterate in bottom-up order through the tree and compute the effective
source terms on each level. This is called the \emph{upward pass} in FMM
literature and computes  $\tilde{f} (\fy_q) = \sum_{j=0}^{M-1} \phi( \fy_j )
f(\fy_j)$ from Eq.\eqref{eq:fmm_conv_combined}.  Second, for each level in the
tree the so-called \emph{level interaction} is computed, where the convolution
of each octant in the level with its influence region is calculated.  This
corresponds to $\tilde{u} (\fx_p) = \sum_{q=0}^{n-1} K(\fx_p, \fy_q)
\tilde{f}(\fy_q)$ in Eq.\eqref{eq:fmm_conv_combined}.  Finally, iterating from
the root to the leaf octants, all contributions are interpolated and
accumulated on the next level, which accounts for  $u (\fx_i) = \sum_{p=0}^{n-1}
\phi(\fx_i) \tilde{u}(\fx_p)$ of Eq.\eqref{eq:fmm_conv_combined}.  This is
called the \emph{downward pass}.
Schematically, the FLGF method  is shown on figure \ref{fig:fmm_pic}
and can be summarized by the following algorithm:

\begin{figure}[!t]
	\centering					
	\includegraphics[width=0.9\textwidth]{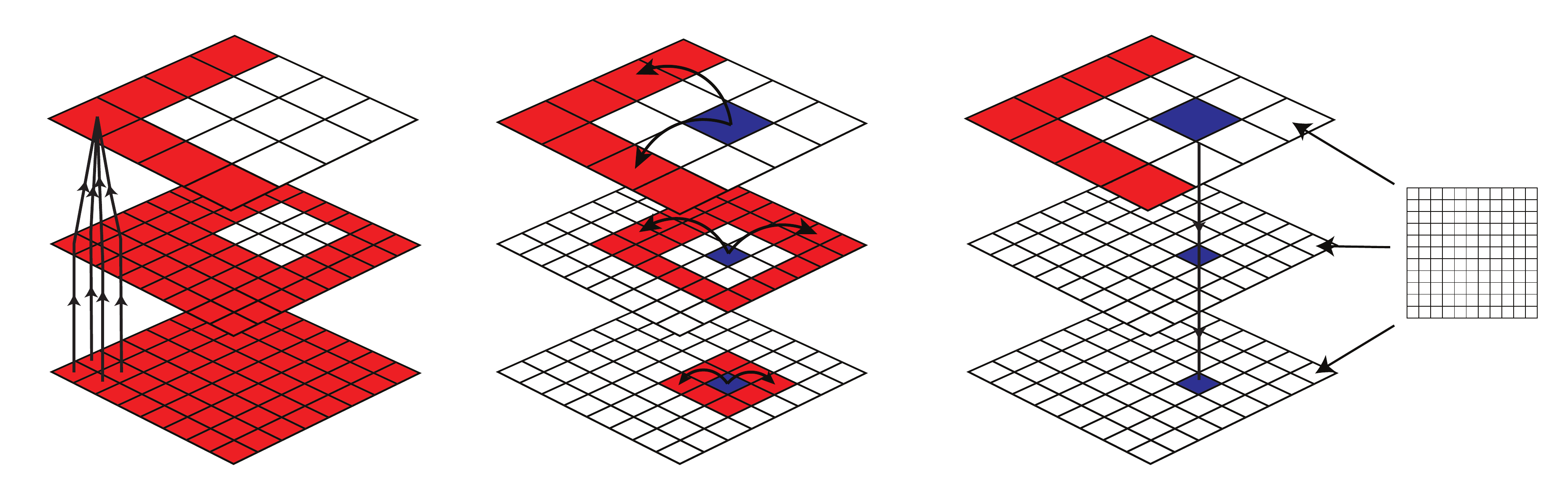}
    \caption{Schematic of the fast multipole method: 
             Left: \emph{Upward pass - }Source regularization . 
             Middle: \emph{Level interaction - }Convolution of a block (blue) with its
             influence list (red). 
             Right: \emph{Downward pass - }Compute and accumulate the induced
             fields at the interpolation nodes. Note that each FMM cell, corresponds
             to a Cartesian block in the FLGF.
             }
	\label{fig:fmm_pic}
\end{figure}


\begin{enumerate}
\item \emph{Upward pass}:
Compute effective source terms at interpolation nodes\\
$\text{For } l=L_{B}-2,...,0: \text{ For } i=0,.. . N_B^l $ \\
\begin{equation}
\tilde{f}_i^l =\sum_{j \in {\mathcal{C}(\mathcal{B}_i^l)}} R^{l+1} \tilde{f}_j,
\end{equation}
where the regularization operator $R^{l+1}$ is the adjoint of the interpolation operator $J^{l+1}$(see below).
\item \emph{Level Interaction }:  FFT Convolution with the octant in the influence region\\
$\text{For } l=0,...,L_{B}-1: \text{ For } i=0,.. . N_B^l $ \\
    \begin{equation}
    \tilde{v}_i^l  =  \sum_{j \in \mathcal{I}(\mathcal{B}_i^l)} {\rm{conv}}(G_{i-j},f_j),
    \end{equation}
where $\rm{conv}(\cdot)$ is the FFT convolution operator.

\item \emph{Downward pass}:  
Compute and accumulate induced field at interpolation nodes \\
$\text{For } l=0,...,L_{B}-1: \text{ For } i=0,.. . N_B^l $ \\
\begin{equation}
\tilde{u}^l_i  = \tilde{v}^l_i  + J_i^{l-1} \tilde{u}^{l-1}_i,
\end{equation}
where the interpolation operator $J^l$ interpolates from the parent onto 
the child block.
\end{enumerate}
Owing to the regularity of the Cartesian block
mesh, the interpolation operators are implemented using Lagrangian polynomials.
As in \cite{liska2014}, we typically use $n_I \leq 10$ interpolation nodes 
to a achieve a relative interpolation error of $\epsilon \approx 10^{-12}$ 
for an analytic function approximation.
The regularization operator is given by the adjoint of the
interpolation operator and is sometimes called anterpolation in FMM literature.
In summary the FLGF method combines the fastest methods for regular meshes,
while retaining the geometrical flexibility and overall linear complexity
inherent to FMM. Excellent computational rates and parallel performance have been
reported in \cite{liska2014}.
In the following, this methodology will be extended to allow for block-refinement.

\section{Block-refined FLGF method}
\label{section:2}
\begin{figure}[!t]
	\centering					
	\includegraphics[width=0.7\textwidth]{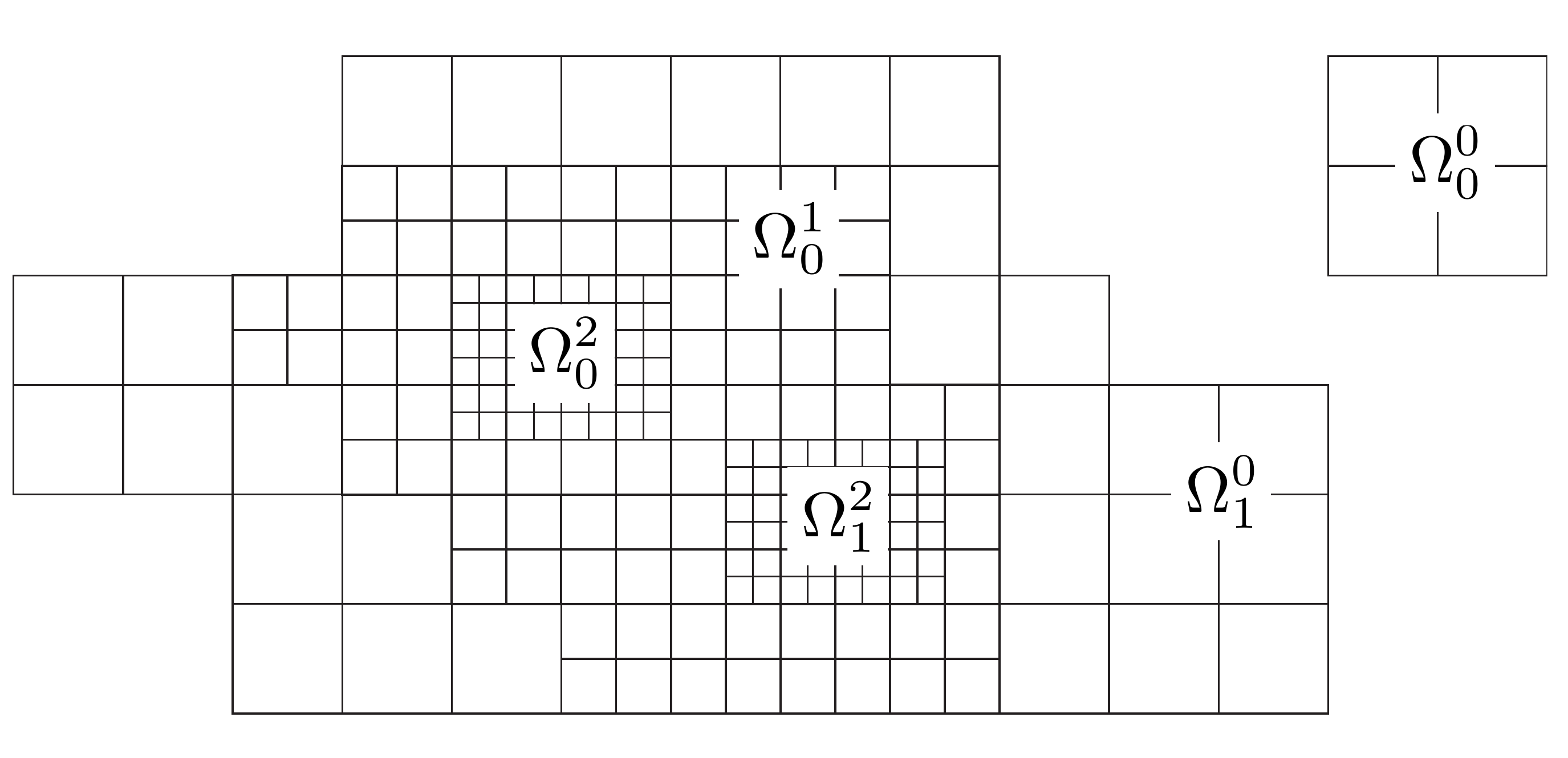}
    \caption{Example of a block-refined mesh topology with two levels of refinement.}
	\label{fig:mesh}
\end{figure}
While the FLGF method has shown to be a fast and promising approach for solving  the
Poisson problem, the methodology is so far limited to
block-structured meshes with uniform resolution and the octree
is only used to compute the block-structured FMM.
Here we propose a multi-resolution, block-refinement strategy and define the computational
domain as $\Omega = \bigcup_{l=0}^{L_{R}-1} \Omega^l  =\bigcup_{l=0}^{L_{R}-1} \bigcup_{m=0}^{M^{\rm{l}}} \Omega_m^{\rm{l}}$,
where $\Omega_m^l$ denotes region $m$ of refinement level 
$l$ and may itself be composed out of $N_m^l$ blocks with $\Omega_m^l = \bigcup_{i=0}^{N_m^l} \mathcal{B}_i^l $
and $\Omega^l= \bigcup_{i=0}^{N_B^l} \mathcal{B}_i^l$ (see figure \ref{fig:mesh}).
In this context, the mesh level $l>=0$ is defined as physical refinement domain.
This is in contrast to the FMM in section \ref{sec:conv}, where the physical domain
was defined on the base level $L_B-1$. This slight abuse of notation will facilitate 
the discussion below.
The refinement method, however can conveniently use the same octree structure
as used for the LGF. We will further use the same nomenclature and distinguish 
between leaf octants and interior, non-leaf octants.

\begin{figure}[!t]
	\centering					
	\includegraphics[width=0.9\textwidth]{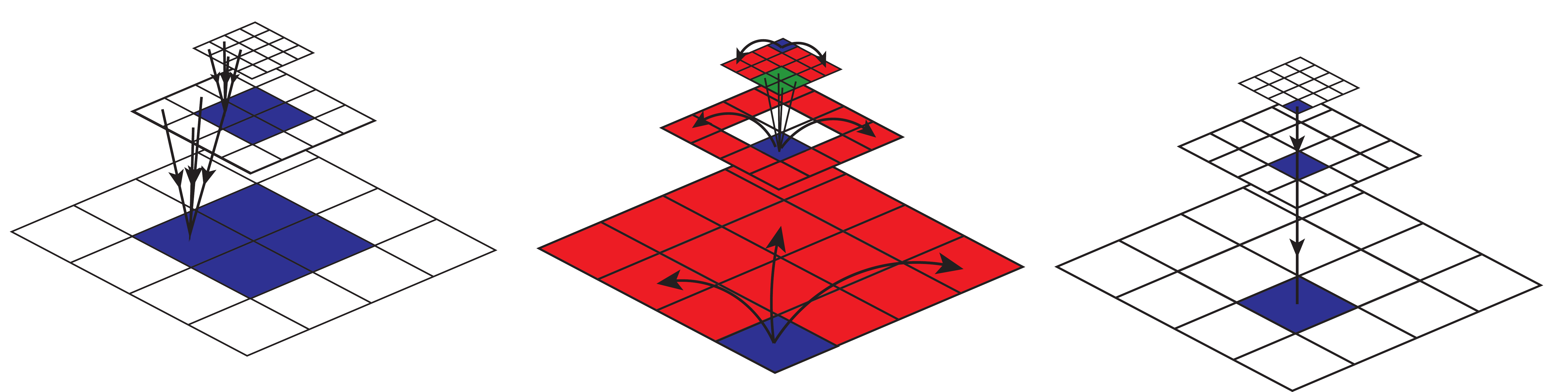}
    \caption{Schematic of the mesh refinement methodology for the FLGF method.
    Left: Coarsification of the source field. 
    Middle: Level interaction (blue-red) and source correction (green).  
    Right: Interpolation and accumulation of the
    target field.}
	\label{fig:amr_all}
\end{figure}

When refining the physical domain by embedding locally refined grid patches
within the computational domain, the free-space boundary conditions implied by
the lattice Green's functions become problematic since the refinement patches
itself do have a well defined boundary condition, which is imposed by the
surrounding domain and is not the free space.
However, if the source field is projected onto each level within its support by
an appropriate coarsening operator, we can, in principle, apply the FLGF method
on each level independently. In terms of the octree structure this corresponds
to recursive coarsening of the source field on all leaf nodes up to the
coarsest refinement level.
We further define the convolution of source regions with the LGF such that each 
target region, which is defined on non-leaf, interior octants only interacts with
leaf octants, whereas leaf octants interact with both leaf and non-leaf octants.
With a subsequent recursive interpolation procedure of the target fields, the
contribution of all octant on all levels to the target region can be
accumulated and accounted for. Schematically, this method is depict in figure 
\ref{fig:amr_all}.
More concisely, we can summarize the scheme by the following
expression for the target field $u_m^l$
\begin{equation}
\label{eq:amr_formular}
u^{l}_m = \Gamma^{l}_m \left[ \sum_{p<l} J^p G^p f^p +
G^{l} \left( \sum_{p \leq l} \prod_{j >p} C^j f^p - \sum_{p<l} L_Q^{p+1} J^p G^p f^p  \right) \right],  
\end{equation}
where  $f^{l} =  f^{l}(\bm{x})$, $\forall \bm{x}  \in  \Omega^{l}$ 
and the projection operator $\Gamma_m^l f= f^l_m $.
In the above, we have also used a shorthand notation for the convolution of the
Green's function $G^l$ with a field $\varphi^l$ on level $l$, which is given 
by
\begin{equation}
G^l \varphi^l = G^l(\fx, \fy) \varphi^l(\fy)=\sum_{\bm{y} \in \Omega^l} G^l(\bm{x} -\bm{y}) \varphi^l(\bm{y}), 
\qquad \bm{x} \in \Omega^l.
\end{equation}
Further, the interpolation from level $l$ onto level $l+1$ is denoted by 
$J^l$, whereas the coarsening operator $C^l$ projects a field from 
level $l$ to level $l-1$.
Note that we have included an additional source
correction term (last term of Eq.~\eqref{eq:amr_formular}).
As will be shown later, this source correction step is not necessary 
to retain second-order accuracy but is used to ensure consistency 
between the Laplace operator $L_Q$ and its inverse.
While this multi-resolution scheme computes the inverse of the multi-resolution
Laplace operator numerically, the corresponding forward Laplace operator is
apriori not known. This is due to the fact that the LGF in Eq.\eqref{eq:GL}
corresponds to the inverse of the 7-pt Laplace operator on a uniform mesh with
free-space boundaries only. 
For practical applications however, it is often necessary to apply at least 
the gradient operator on the target field. 
Hence, it is desirable to know or at least be able to apply a consistent
forward operator and we seek Laplacian free field for each level, where the
following equation to holds:
\begin{equation}
L_Q^l \Gamma^l  \sum_{p<l} J^l G^l f^l=0, \quad l =0,.. .,L_{R}-1.
\end{equation}
The source correction step ensures consistency between the forward operator per level, in
this case the 7-pt Laplace, and the computed inverse by numerically evaluating 
the error, originating from said inconsistency,
on the next level and subtracting it from the source on this level.
This is indicated in figure \ref{fig:amr_all}b for the second and third
refinement level with the green region. Naturally, this is only applicable for
non-leaf octants.

Using the notation from section \ref{sec:conv}, the algorithm for a target field $u^l_i$,
defined on octant $\mathcal{B}_i^l$, is given by the following steps:
\begin{enumerate}
\item  Regularization:\\ 
$\text{For } l=L_{R}-2,...,0: \text{ For } i=0,.. . N_B^l $ \\
       \begin{equation}
        \tilde{f}^l_i =\sum_{ j\in C(\mathcal{B}_i^l)} C^{l+1}f_{j}
       \end{equation}

\item  Level interaction:\\ 
$\text{For } l=0,...,L_{R}-1: \text{ For } i=0,.. . N_B^l $ \\
       \begin{enumerate}
       \item Convolution:
       \begin{equation}
       \label{eq:amr_convs}
           \tilde{v}_i^l=
           \begin{dcases}
           \sum_{j \in B^l} {\rm{conv}}(G_{i-j}, f^l_j)  \quad &\text{if } \mathcal{B}_i^l \in B^l_{\rm{Leafs}} \\
           \sum_{j \in B^l_{\rm{Leafs}}} {\rm{conv}}(G_{i-j}, f^l_j)  \quad &\text{else}.
           \end{dcases}
       \end{equation}
       \item Source correction:
       \begin{equation}
       \tilde{f}^{l+1}_i \leftarrow \tilde{f}_i^{l+1} - L_{\mathcal{Q}}^{l+1} J^l \tilde{f}_i^l
       \end{equation}
       \end{enumerate}
\item  Accumulation and interpolation:\\ 
$\text{For } l=0,...,L_{R}-1: \text{ For } i=0,.. . N_B^l $ \\
    \begin{equation}
    \tilde{u}^{l}_i = \tilde{v}^{l}_i + J^{l-1} \tilde{u}^{l-1}_i
    \end{equation}
\end{enumerate}

%
Note that the convolution in the second step is computed using the FLGF method 
as presented in section \ref{sec:conv}.
Since, we need to distinguish between the interaction of leaf nodes
and the interior octants, we also have the carry out the FMM twice. 
In practice, we first use the FMM for computing the convolution of the entire 
refinement tree and a second time using the non-leafs nodes only. 
Subtracting both is equivalent to Eq.\eqref{eq:amr_convs} and yields
\begin{equation}
\tilde{v}^l(\fx) = G^l(\fx, \fy) f^l(\fx) - G^l(\fx, \fy_I) f(\fy_I) \qquad \forall \fx \in B^l, \fy_I \in B^l \setminus B^l_{\rm{Leafs}}.
\end{equation}
Note that the computational complexity remains linear in the number DoFs per level 
and also scales linearly with the number of refinement levels and thus retains
linear complexity in the total number of DoFs.

\subsubsection*{Regularization and Interpolation operators}
\begin{figure}[!b]
	\centering					
	\includegraphics[width=0.45\textwidth]{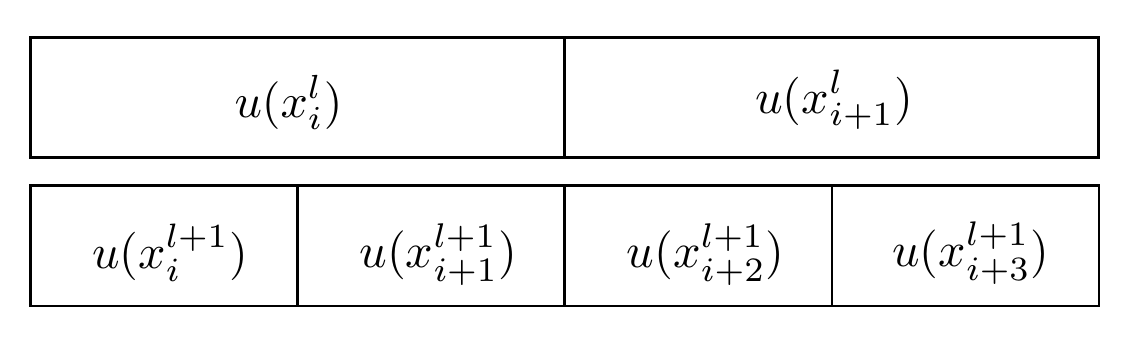}
    \caption{Mesh refinement in one dimension.}
	\label{fig:mesh_1d}
\end{figure}

The FLGF method is a second-order accurate scheme and we thus require 
the interpolation and regularization operators to be at least second-order
accurate to avoid degradation of the overall accuracy of the scheme.
The field values are stored at the cell centers, which leads to 
a staggered storage upon refinement, 
i.e., none of the field values in any child cell are stored in the same
physical location than the field values of the parent cell (see figure \ref{fig:mesh_1d}). 
This is in contrast to 
vertex storage, which would lead to field values stored at coinciding locations
for different levels. 
For ease of implementation, we define the interpolation operator to be the
second-order Lagrangian polynomials, which already have been used for the FMM.
Note that in the procedure above, the interpolant corresponds to the
interaction of a non-leaf octant with the leaf octants only. Thus the fields
within each non-leaf octant are distinct.  We use an additional buffer layer of
one cell around each octant and set the source within this region to zero. This
allows us to use an unbiased interpolation for each octant.
\\
For the coarsening operator, we use a simple averaging procedure, which is
second-order accurate and does not require any neighbor information.  In one
dimension this yields 
\begin{equation}
u(x_i^{l}) = { u(x_i^{l+1})+ u(x_{i+1}^{l+1}) \over 2}.
\end{equation}

\subsection{Convergence}

\begin{figure}[!t]
    \centering
	\begin{subfigure}[b]{0.35\textwidth}
        \includegraphics[width=\textwidth]{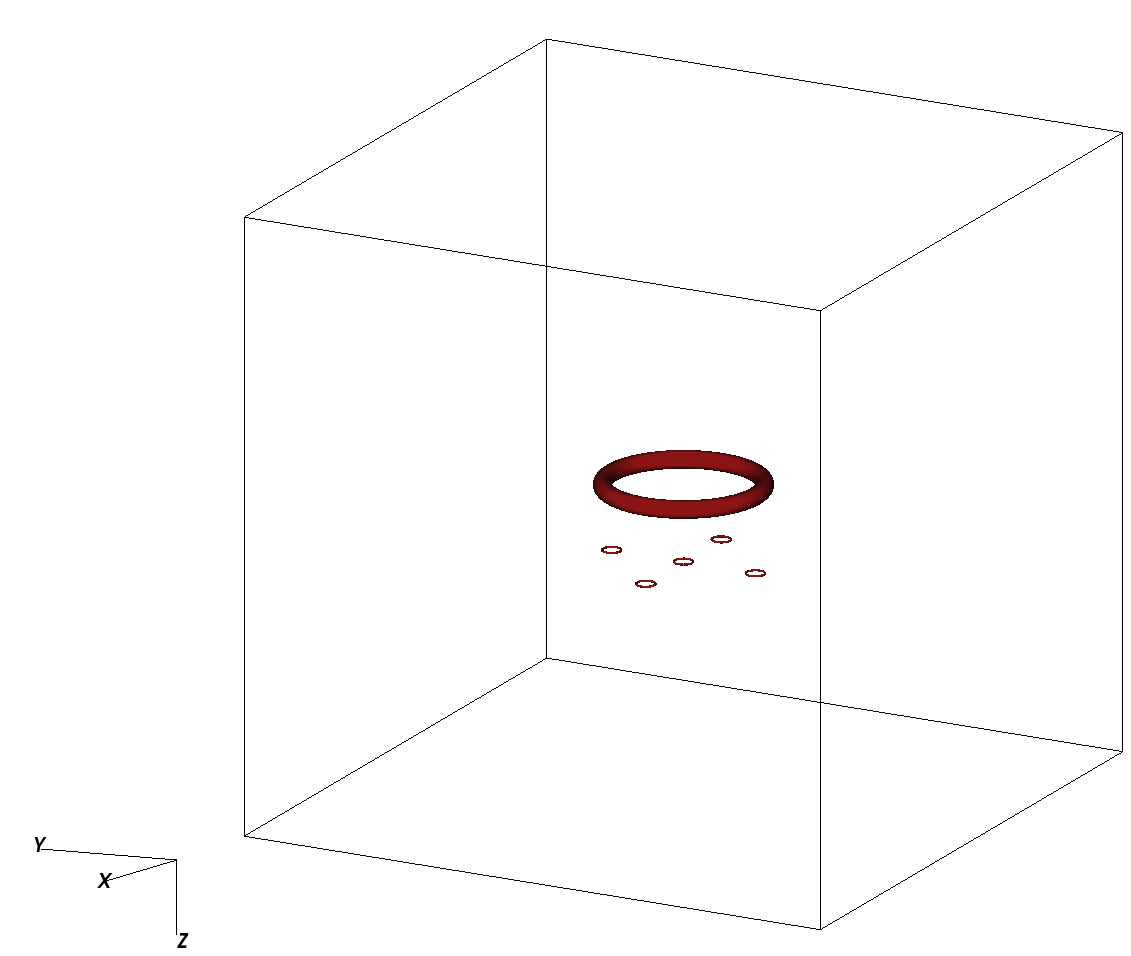}
    \caption{Arrangement of the vortex rings.}
    \label{fig:setup_arrangment}
	\end{subfigure}
	\begin{subfigure}[b]{0.64\textwidth}
        \includegraphics[width=\textwidth]{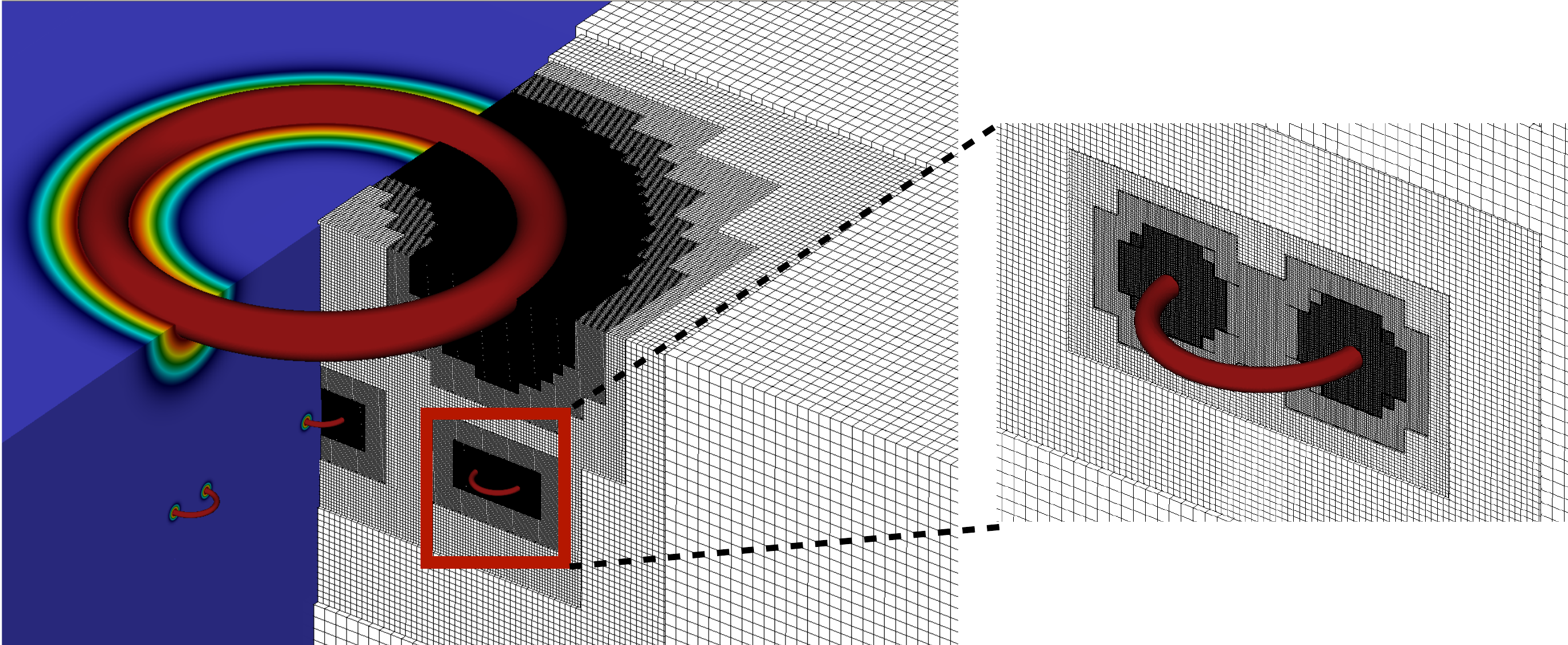}
    \caption{Mesh topology for the vortex rings and six levels of refinement.}
    \label{fig:setup_mesh}
	\end{subfigure}
    \caption{Mesh topology and vorticity field for the vortex ring and
    the numerical solution of the streamfunction $\Psi_h$
             for six levels of refinement.}
\end{figure}

For validation of the multi-resolution scheme as presented above, 
we use the method of manufactured solutions. As a test problem we consider
a vortex ring with radius $R$ and its streamfunction $\Psi$ is defined as 
\begin{equation}
    \label{eq:streamfunc}
    \Psi(r,z)=f \left( \frac{\sqrt{(r-R)^2+z^2}}{R}   \right)\bm{e}_{\theta}, 
\end{equation}
where 
\begin{equation}
    \label{eq:streamfunc_f}
    f(t) = 
    \begin{dcases}
    c_1 \exp \left( - \frac{c_2}{1-t^2}\right) \quad \text{if } &|t|<1 \\
    0.0 & \text{else}.
    \end{dcases}
\end{equation}
The streamfunction is related to vorticity by the Poisson equation $\omega= \triangle \Psi$.
This allows us to initialize the source field by analytical evaluation of the 
Laplace and compare the numerically obtained solution of the streamfunction 
with the analytical one in Eq.\eqref{eq:streamfunc}.
In this setup we place six vortex rings in the unit cube, 
where a single large vortex ring with $c_1=10^{3}$, $c_2=10$ and radius $R=0.125$ 
is located in the center of the domain and five smaller once's with $c_1=10^{6}$, $c_2=15$ and radius $R=0.015$ at $z=0.125$ 
are arranged as shown in figure \ref{fig:setup_arrangment}.

As a first convergence test, a global mesh refinement study with a fixed 
mesh topology of three levels near the vortex ring is carried out. 
An example mesh topology with six levels of refinement is shown in figure \ref{fig:setup_mesh}.
The mesh is composed of cubic blocks and we initialize or refine the block if on level $l$ the following 
criterion is met
\begin{equation}
    \omega^l >  \alpha^{L_R-l}\omega^l_{{\rm{max}}},
    \label{eq:ref_crit}
\end{equation}
where $\alpha=1/32$ unless stated otherwise.
This criterion is conservative and is offered for the specific test problem
shown; it should be reevaluated for different applications.

\begin{figure}[!t]
	\centering					
	\includegraphics[width=0.6\textwidth]{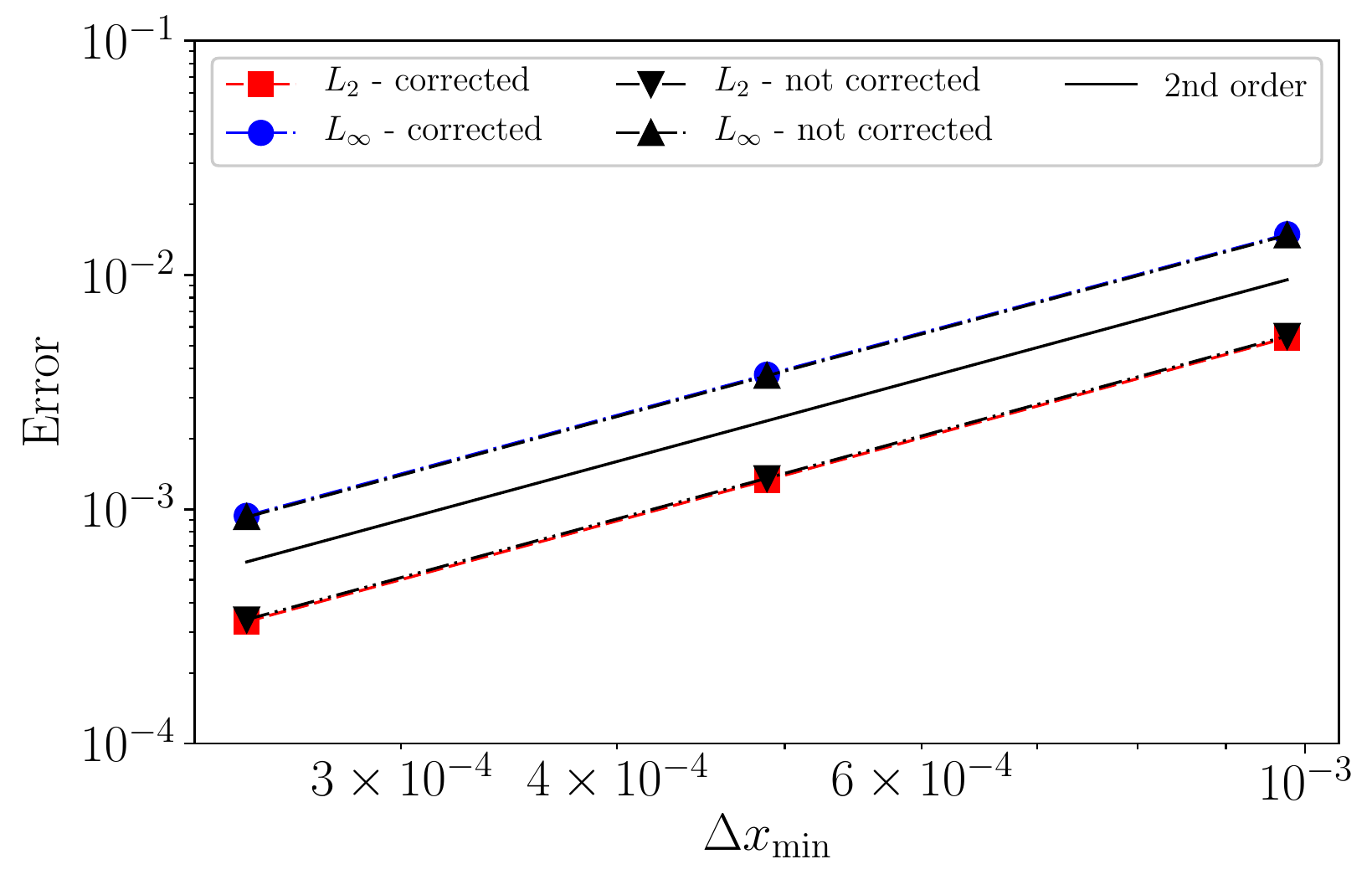}
    \caption{Convergence of the $L_2$ and $L_\infty$ norm with respect to the
    analytical target field $\Psi$ for the block-refined FLGF method for a
    fixed mesh topology of three refinement levels and an increasing base level
    resolution $\Delta x_{\rm{max}}$.}
    \label{fig:convergence_amr}
\end{figure}

In figure \ref{fig:convergence_amr}, the $L_2$ and $L_\infty$ norm with respect to the analytical solution 
are shown with increasing base level resolution. As expected a clear second-order convergence behavior 
is observed for the proposed multi-resolution scheme. We also plot the target error for the case, where
the source correction term is neglected. From the figure it is apparent that the influence of the source
correction term on both the convergence and the magnitude of target error is
negligible for global refinement and a total of three refinement levels.

\begin{figure}[!t]
    \centering
	\begin{subfigure}[t]{0.49\textwidth}
	\includegraphics[width=\textwidth]{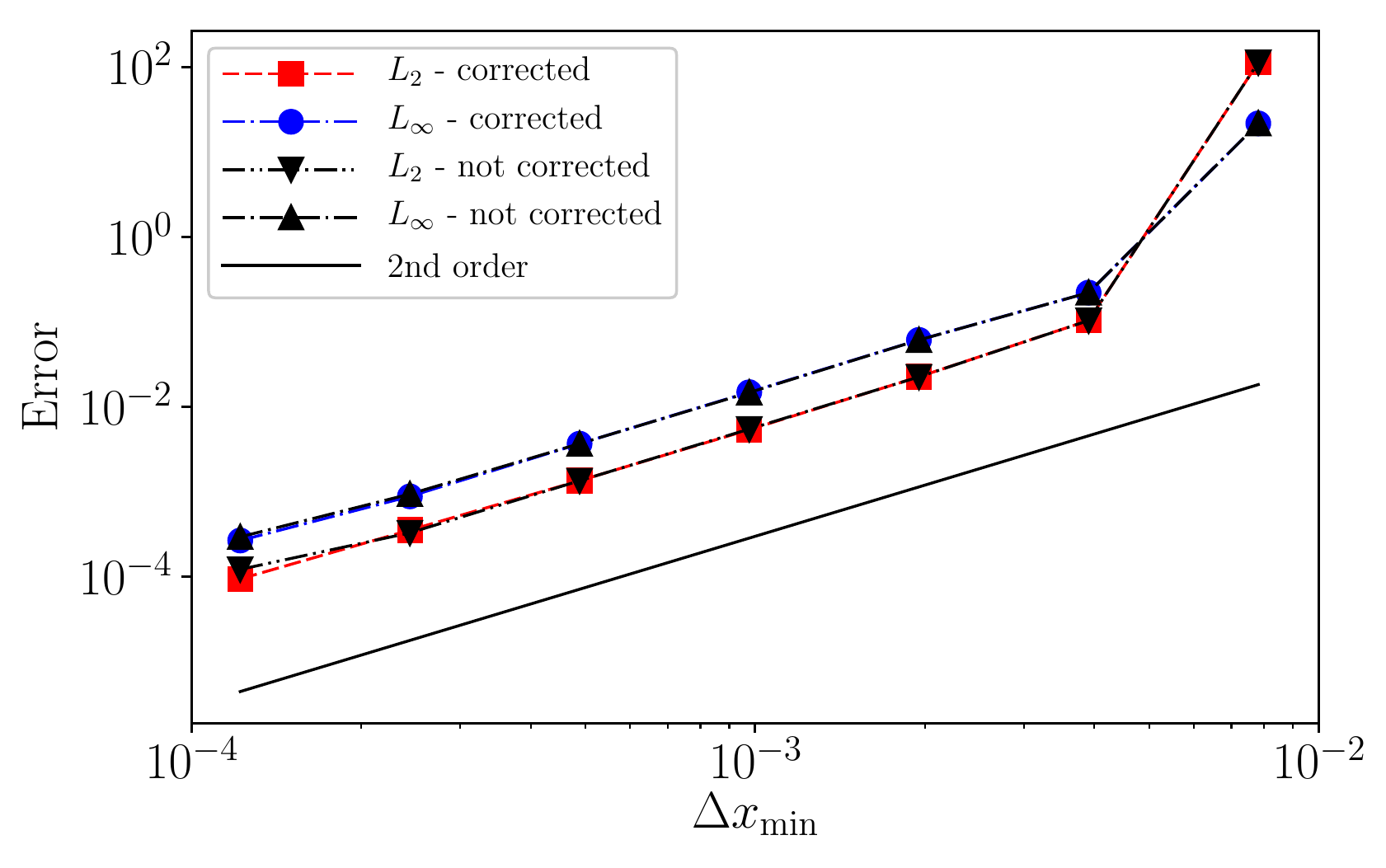}
    \caption{$L_2$ and $L_\infty$ norm in the finest level with respect to the
    analytical target field $\Psi$ for increasing levels of refinement.}
    \label{fig:TargetError_incLevels}
	\end{subfigure}
	\begin{subfigure}[t]{0.49\textwidth}
	\includegraphics[width=\textwidth]{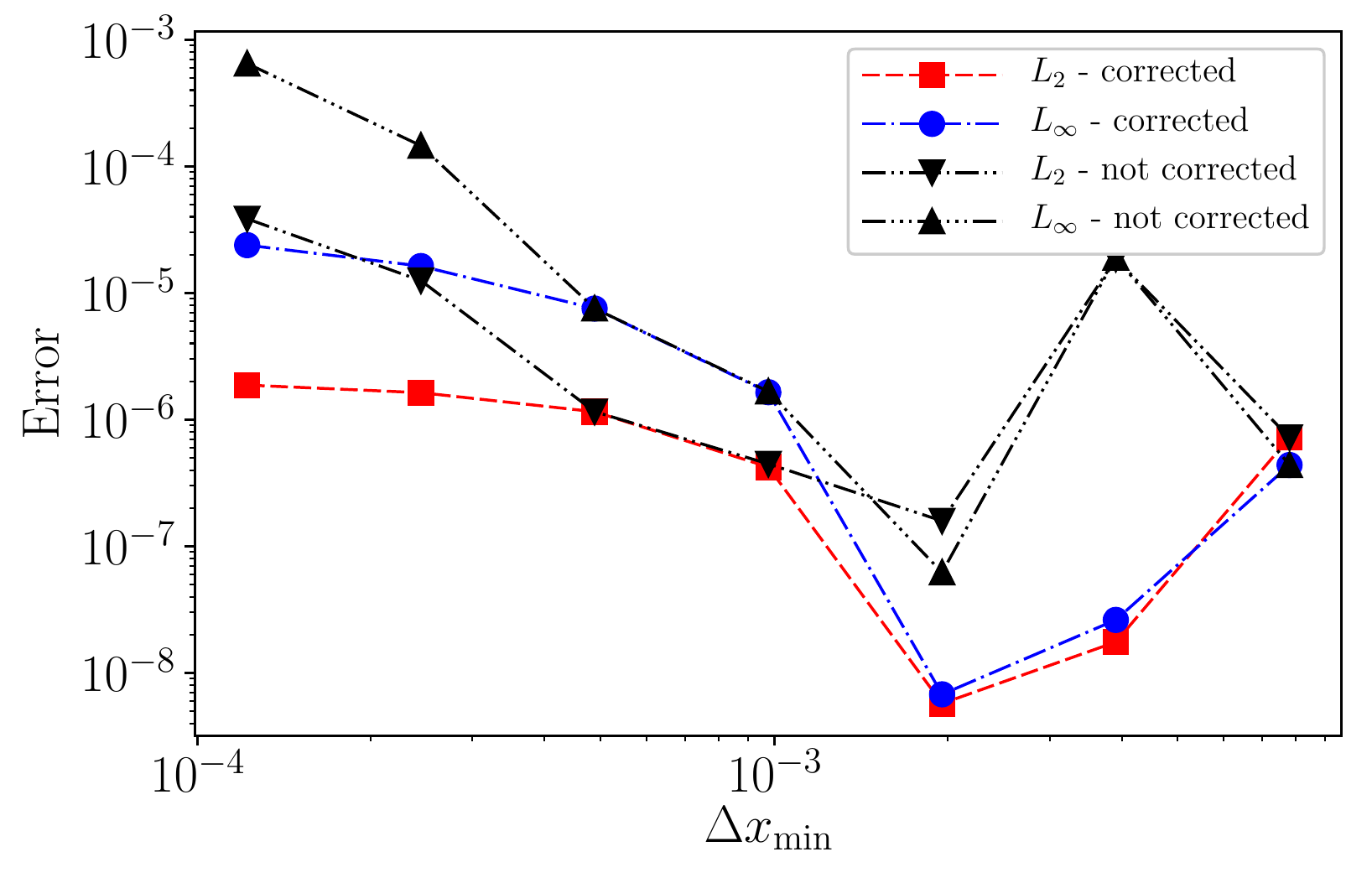}
    \caption{$L_2$ and $L_\infty$ norm of $\omega_h=L_\mathcal{Q} \Psi_h$ with
    respect to the analytical source field $\omega$ for increasing levels of
    refinement.}
    \label{fig:offset_target_convergence}
	\end{subfigure}
    \caption{Convergence study of six vortex rings with refinement
    coefficient $\alpha=1/32$}
    \label{fig:convergence_w32}
\end{figure}

\begin{figure}[!t]
    \centering
	\begin{subfigure}[t]{0.49\textwidth}
	\includegraphics[width=\textwidth]{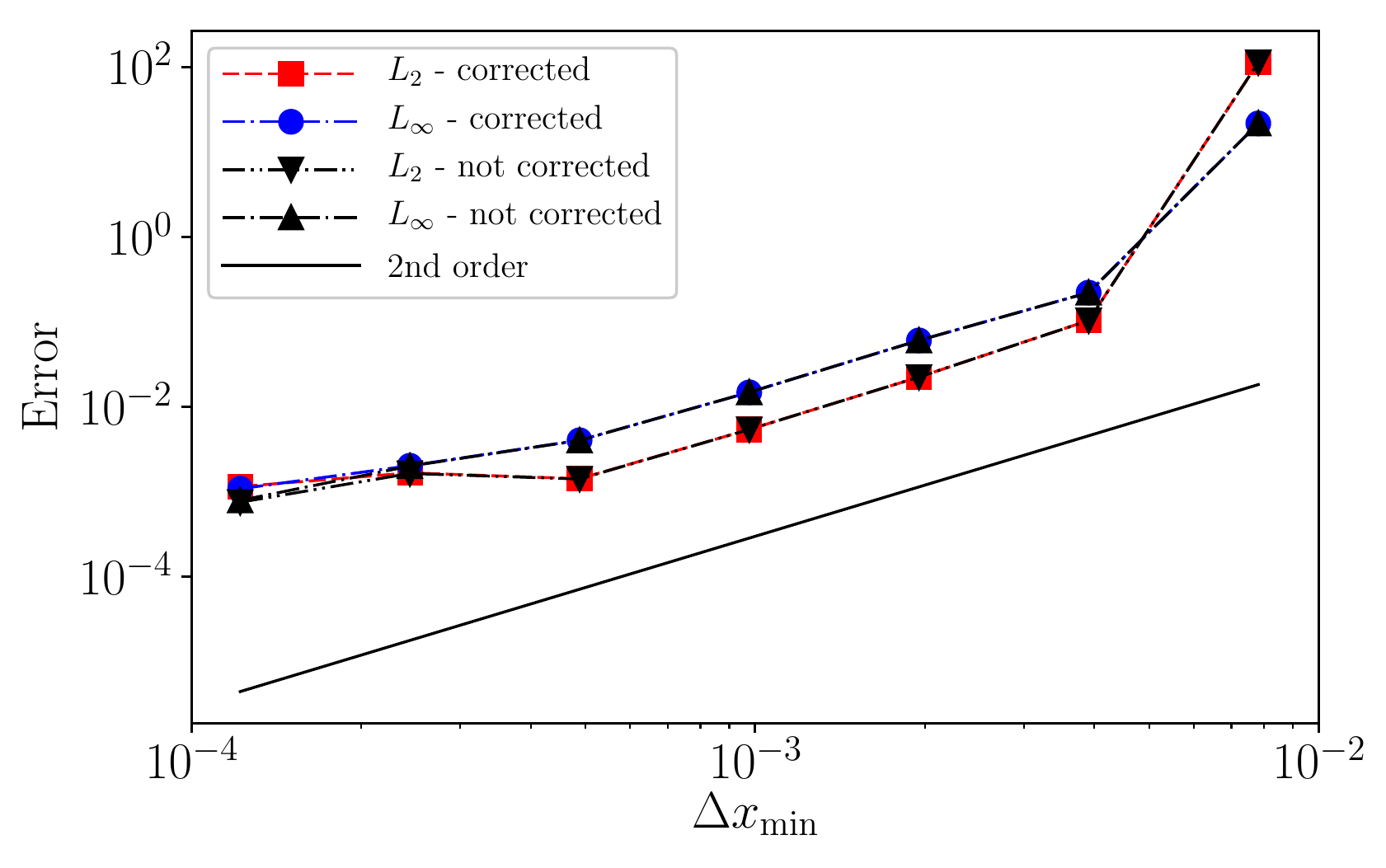}
    \caption{$L_2$ and $L_\infty$ norm in the finest level with respect to the analytical target field $\Psi$
            for increasing levels of refinement.}
    \label{fig:TargetError_incLevels_w8}
	\end{subfigure}
	\begin{subfigure}[t]{0.49\textwidth}
	\includegraphics[width=\textwidth]{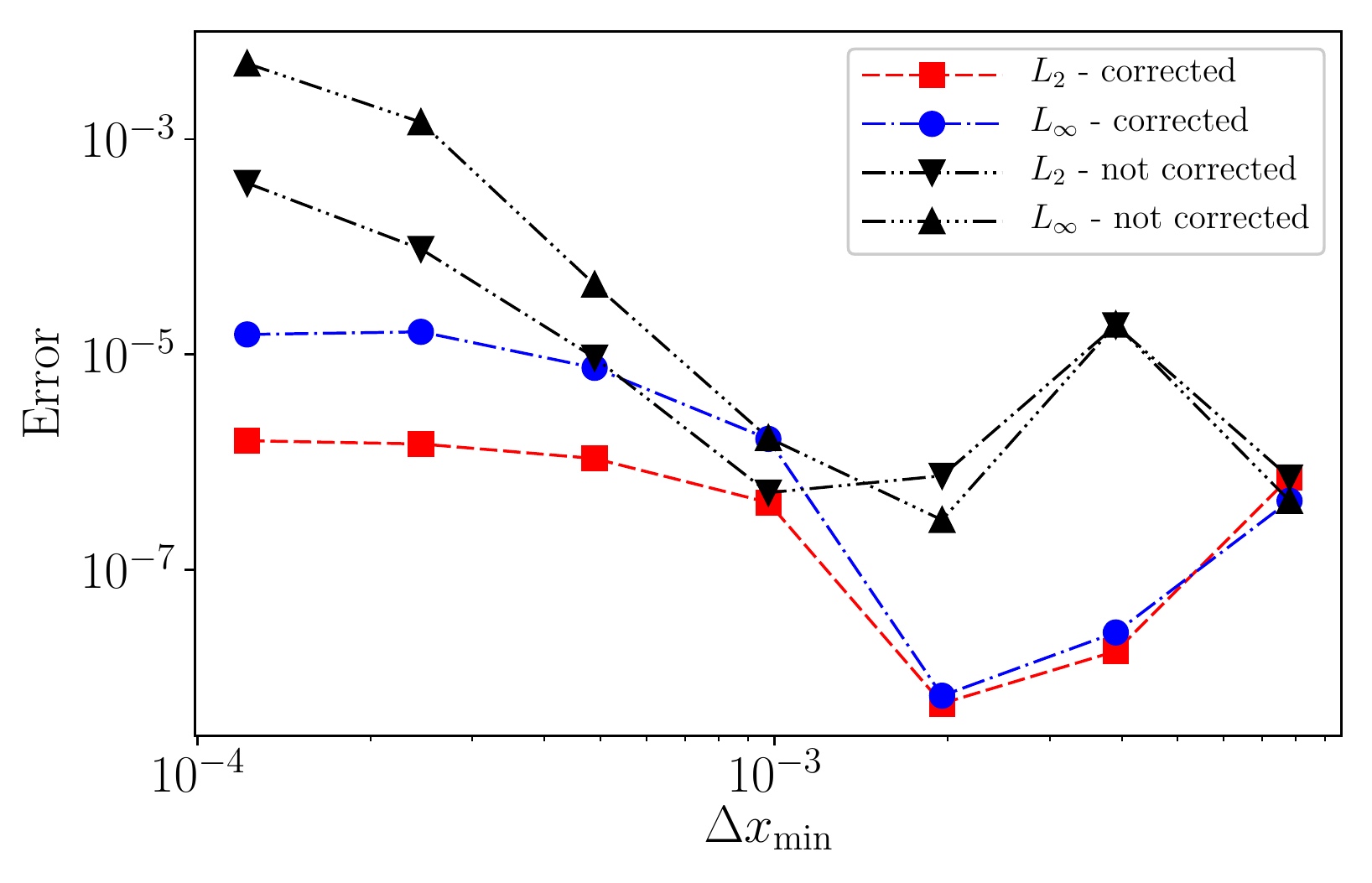}
    \caption{$L_2$ and $L_\infty$ norm of $\omega_h=L_\mathcal{Q} \Psi_h$ with respect to the analytical source field $\omega$
            for increasing levels of refinement.}
    \label{fig:offset_target_convergence_w8}
	\end{subfigure}
    \caption{Convergence study of six vortex rings with refinement
    coefficient $\alpha=1/8$}
    \label{fig:convergence_w8}
\end{figure}

%
%
%
In addition to changing the base level resolution, we now fix the base level 
resolution and increase the number of refinement levels successively according
to the refinement criterion in Eq.\eqref{eq:ref_crit}.
The corresponding convergence of the error in the fine level for both the
corrected and the uncorrected version is shown in figure
\ref{fig:TargetError_incLevels}. 
Note that while second-order of accuracy is always retained in the proposed
scheme when refining a fixed mesh topology on all refinement levels uniformly
(see figure \ref{fig:convergence_amr}), the order of accuracy will degrade if
non-negligible source is contained in non-refined regions as apparent in figure
\ref{fig:TargetError_incLevels}. The saturation of the error is dependent on
the specific refinement criteria chosen
as exemplary shown in figure \ref{fig:TargetError_incLevels_w8},
where the refinement criterion was relaxed and $\alpha=1/8$ was used.


We also investigate the effect of the source correction term on the consistency
between forward Laplace operator $L_\mathcal{Q}$ and its numerically computed
inverse $\Psi_h$. To that end, we take the 7-pt Laplacian of the target field
for each level and compare it to the source field. On a uniform mesh, the
source field is recovered up to the precision of the FMM. For the multi-resolution
mesh, the errors are reported in figure
\ref{fig:offset_target_convergence}.  The figure shows, as expected, that
the error of the source field $\omega_h=L_\mathcal{Q} \Psi_h$ compared to
$\omega$ becomes large and increases with the number of refinement levels
when the appropriate source correction terms are not included.  On the
other hand, the correction has a relatively minor impact on the  target
field itself, $\Psi_h$, as shown in figure~\ref{fig:TargetError_incLevels},
but ensures consistency between the forward Laplacian and its inverse.
Analogous behavior is observed in figure \ref{fig:offset_target_convergence_w8} for
the case of $\alpha=1/8$.



\section{Efficiency and parallel performance}
To demonstrate the efficiency and parallel performance of the propose 
block-refined algorithm, we consider the same test case as above and report 
its efficiency and parallel performance. 
We also refer to \cite{liska2014} for performance investigations and 
the parallel implementation strategy of the uniform solver, which  
is similar to the one employed here.
However, for completeness we also report the computational rates here. 
\begin{figure}[!t]
	\centering
	\begin{subfigure}[t]{0.49\textwidth}
	\includegraphics[width=\textwidth]{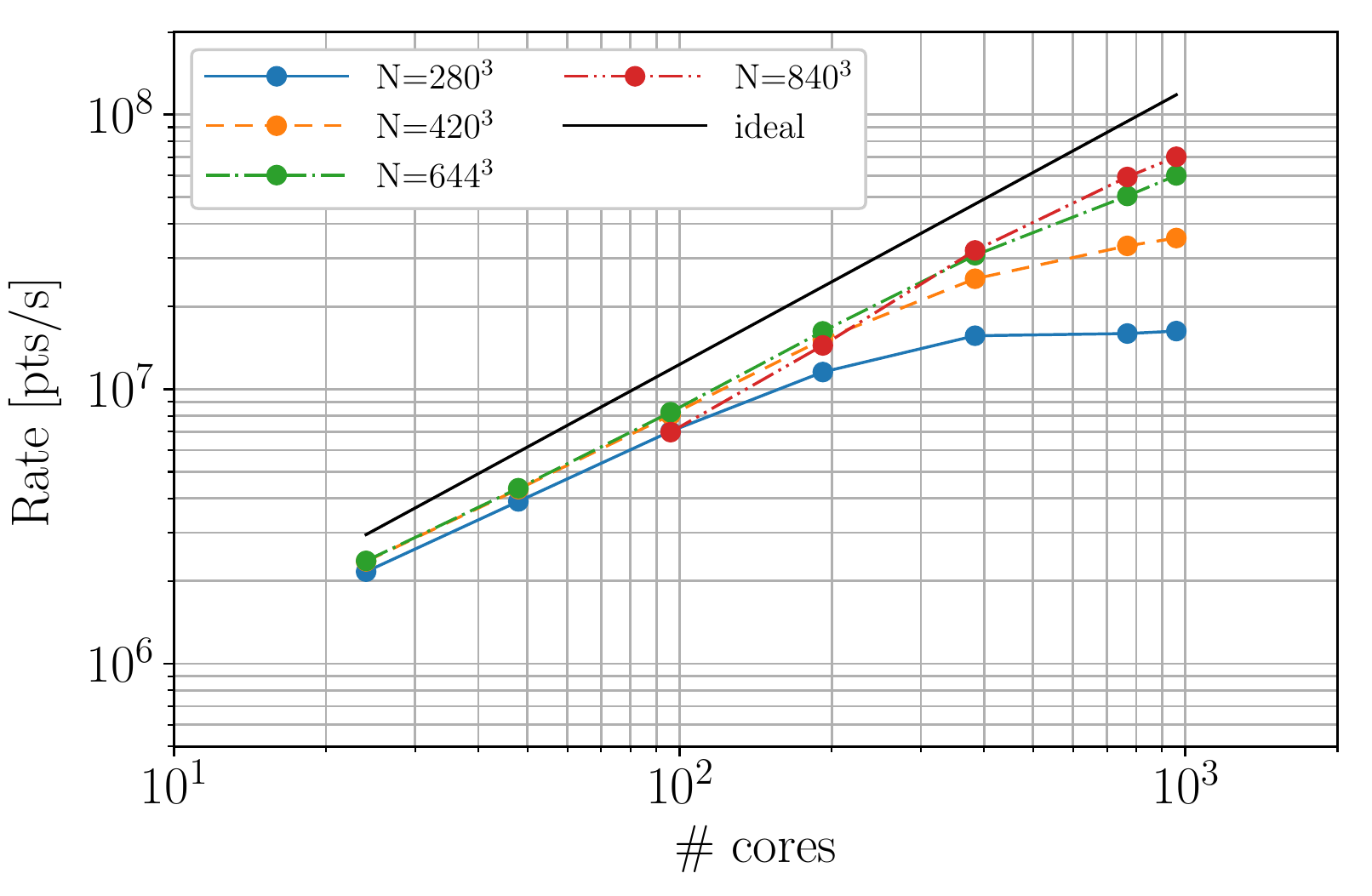}
    \caption{Computational rates for a uniform mesh of size $N$. }
    \label{fig:level0_eff}
 	\end{subfigure}
	\begin{subfigure}[t]{0.49\textwidth}
	\includegraphics[width=\textwidth]{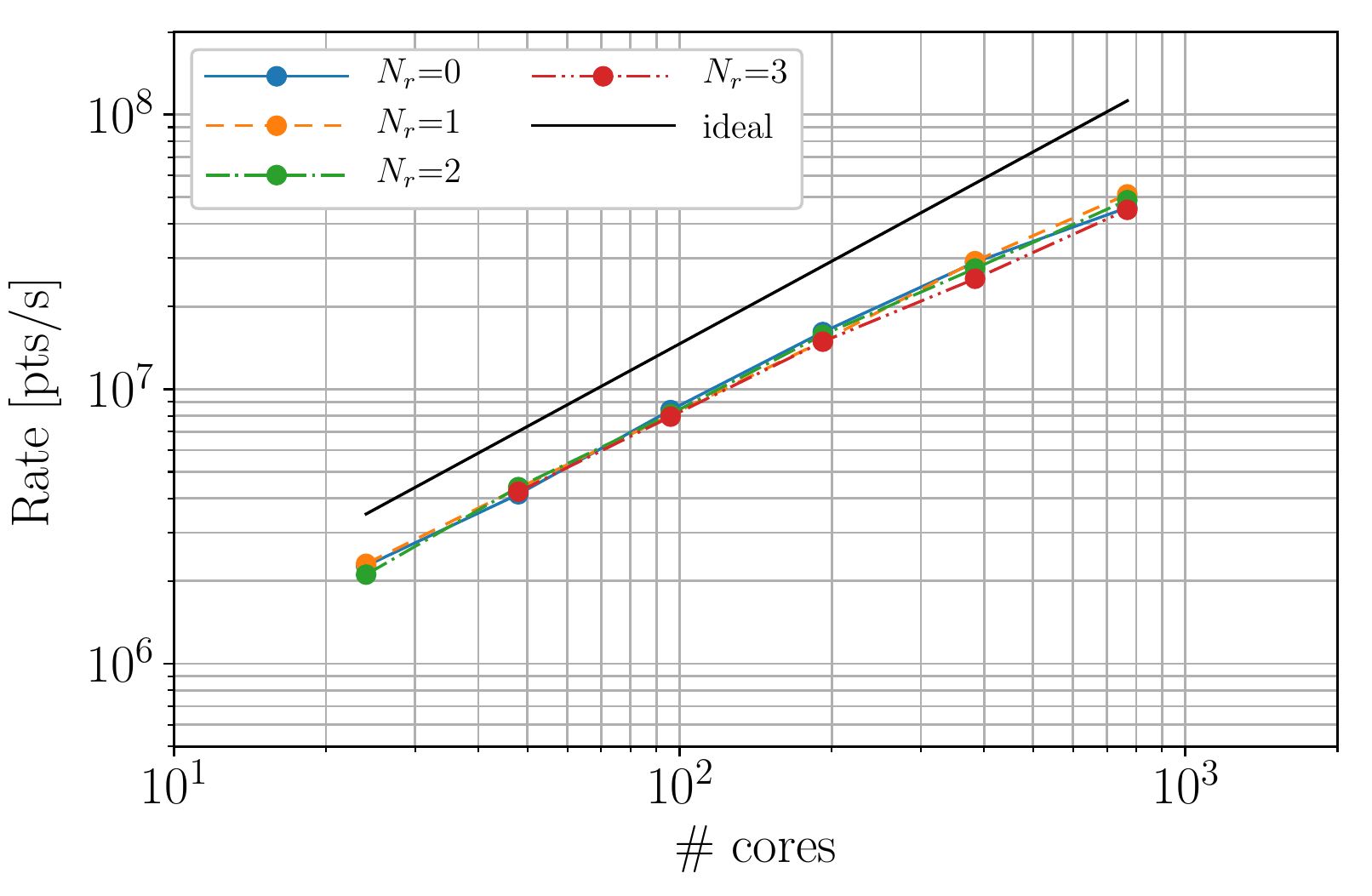}
    \caption{Computational rates for $N=644$ and an increasing number of refinement levels.}
    \label{fig:level_eff}
 	\end{subfigure}
    \caption{Computational rates and parallel performance. }
    \label{fig:efficiency}
\end{figure}

The solver was written in a C++ framework and uses MPI for parallel
communications as well as FFTW for fast Fourier transforms.  The code
implements an octree data structure, where each leaf corresponds to a cubic
domain in physical space.  A server-client model is used for load balancing of
the octree, where the sever stores the full octree but does not allocate any
data. The clients on the other hand store one or multiple sub-trees including
the corresponding data. For load balancing, the anticipated load (mainly
the load of fast Fourier transforms) is computed for each octant and the leaf
octants are sorted according to their Morton code for each level. Finally, the
sorted array of leaf octants is split into junks with almost equal loads, which
are then assigned to each processor.  Subsequently, the parents are assigned to
the processor with the minimum load in a recursive fashion.  Note that the FMMs
are sequential in terms of the level due to the correction term, which depends
on previous levels and thus necessitates level-wise balancing to avoid an
imbalance of load on a particular level.  All communication patterns between
clients are established using the server and communication costs are almost
fully hidden using non-blocking MPI calls.
\\
Note that by far most time is spend in the level interactions ($\sim 99\%$) of the algorithm
and in particular the level convolution within each FMM. 
The time to construct or traverse the octree data structure is negligible due to 
the block-wise nature of the algorithm, which also allows SIMD vectorization
of the Fourier transforms and the Hadamard product for additional speed.
\\
Figure \ref{fig:level0_eff} shows the strong scaling for the computational
rates of our implementation for various domain sizes $N$. The parallel efficiencies
are in line with the implementation of \cite{liska2014} as well as other
kernel-independent FMM solvers and thus verifies our implementation.
In addition, in figure \ref{fig:level_eff}, the dependence of the computational 
rate with the number of refinement levels is plotted for the case of $N=644^3$.
As stated earlier, the complexity scales linearly with the number of levels 
and we thus expect that the computational rates are independent of the  refinement 
levels. This is confirmed in \ref{fig:level_eff}. 
Note however that the parallel efficiency is limited to the efficiency of the 
individual FMMs.  

\section{Discussion and Conclusion}
In this paper, we presented a multi-resolution scheme for inhomogeneous,
linear, constant-coefficient difference equations on infinite grids based on an
extension of the fast lattice Green's function (FLGF) method.
The new method retains inherent advantages of the FLGF
method, such as the use of uniform mesh interpolations and FFT-based
convolutions, and the inherent computational savings associated with
satisfaction of the exact far-field boundary conditions for domains truncated
snugly around source regions, but allows for an arbitrary block-wise mesh
refinement by factors of two.  The refinement procedure consists of
regularization, level interaction as well as interpolation. By consistent
definition of all operators, second-order accuracy is retained, and the
complexity of the scheme remains linear in the number of DoFs.
This was demonstrated for different refinement criteria using manufactured
solutions for the Poisson equation of the streamfunction of multiple vortex
rings at different spatial scales. Linear complexity in the number of
refinement levels as well as parallel efficiency was shown numerically. Hence,
the method is well suited for large-scale computation of  multi-scale phenomena
that render conventional solvers based on uniform grids or nonlinear complexity
impractical.


\section*{Acknowledgments}
This work was supported by the SNF Grant No. P2EZP2\_178436 (B. D.)  and the ONR Grant No. N00014-16-1-2734.

\bibliographystyle{model1-num-names}
\bibliography{references.bib}

\end{document}